\documentclass[aps,prb,twocolumn,superscriptaddress,showpacs]{revtex4}
\usepackage{amsmath}
\usepackage{dcolumn}
\usepackage{amssymb}
\usepackage{amsfonts}
\usepackage{graphicx}
\usepackage[T1]{fontenc}
\usepackage{indentfirst}

\begin{document}
\title{Interaction-induced anomalous quantum Hall state on the honeycomb lattice}
\author{Tanja \DJ uri\'c}
\affiliation{London Centre for Nanotechnology, University College London, 17-19 Gordon St, London, WC1H 0AH, UK}
\author{Nicholas Chancellor}
\affiliation{London Centre for Nanotechnology, University College London, 17-19 Gordon St, London, WC1H 0AH, UK}
\author{Igor F. Herbut}
\affiliation{Department of Physics, Simon Fraser University, Burnaby, British Columbia, V5A 1S6, Canada}
\affiliation{Max-Planck-Institut f\"{u}r Physik Komplexer Systeme, N\"{o}thnitzer Str. 38, 01187 Dresden, Germany}

\date{\today}
\begin{abstract}

We examine the existence of the interaction-generated quantum anomalous Hall phase on the honeycomb lattice. For the spinless model at half filling, the existence of a quantum anomalous Hall phase (Chern insulator phase) has been predicted using mean-field methods. However, recent exact diagonalization studies for small clusters with periodic boundary
condition have not found a clear sign of an interaction-driven Chern insulator phase. We use exact diagonalization method to study properties of small clusters with open boundary condition and, contrary to previous studies, we find clear signatures 
of the topological phase transition for finite size clusters. We also examine applicability of the entangled-plaquette state (correlator-product state) ansatz 
to describe the ground states of the system. Within this approach the lattice is covered with plaquettes and the ground state
wave-function is written in terms of the plaquette coefficients. Configurational weights can then be optimized using a variational Monte Carlo algorithm. 
Using the entangled-plaquette state ansatz we study the ground state properties of the system for larger system sizes and show that the results agree with the 
exact diagonalization results for small clusters. This confirms validity of the entangled-plaquette state ansatz to describe the ground states of the system and provides
further confirmation of the existence of the quantum anomalous Hall phase in the thermodynamic limit, as predicted by the mean-field theory calculations.

\end{abstract} 
\maketitle
\section{Introduction}
\label{sec:intro}
\indent
In recent years, the interest in topological phases of matter and topological phase transitions\cite{Haldane, Haldane2, Wen} was renewed by the discovery of topological
insulators and superconductors.\cite{Qi, Hasan,Kane1, Kane2, Bernevig, Konig} For noninteracting insulators possible topological states have been classified\cite{Kitaev, Schnyder} 
and the transitions between different topological states have been found to be characterized by closing of the single-particle gap.\cite{Haldane, Qi, Hasan} Much of the recent research
has focused on understanding the role of interactions in topological insulators.\cite{Hohenadler} Of particular interest are interaction-generated topological insulators,
or topological Mott insulators.\cite{Raghu, Weeks, Grushin, Budich} Such states appear due to the presence of electronic interactions that give rise to nonlocal complex bond order 
parameters and spontaneous breaking of time reversal symmetry.\cite{Raghu, Weeks, Grushin,Wu} The possibility of realizing novel interaction-generated topologically nontrivial
phases, without requirement of strong intrinsic spin-orbit coupling, could significantely extend the class of topologically nontrivial materials and is thus of great 
importance. 
\\
\indent
Interacting spinless fermions on a honeycomb lattice is perhaps the best studied example in which an interaction-generated topologically nontrivial state, a quantum anomalous
Hall (QAH) state, is predicted to appear by several mean-field theory calculations.\cite{Raghu, Weeks, Grushin} However, recent exact diagonalization (ED) studies of small clusters with 
periodic boundary condition\cite{Daghofer, Martinez} find no clear sign of a such topologically nontrivial state. In this manuscript we present ED results
for small clusters with open boundary condition. Contrary to previous ED studies,\cite{Daghofer, Martinez} we find clear signatures of the interaction-driven topological transition in the fidelity
metric\cite{Varney, Wang, Gu, Abasto, Eriksson, Varney2, Zanardi,Venuti, Rigol, Varney3} and show that the transition is characterized by closing of the excitation gap. We also demonstrate the existence of the edge states in the QAH phase by 
calculating the density profile of a hole created out of the ground state at half-filling.\cite{Li, Varney4}
\\
\indent
To study properties of the system for larger system sizes we use entangled plaquette state (EPS) ansatz, also called correlator product state (CPS) ansatz,
\cite{Changlani,Mezzacapo1,Neuscamman,Mezzacapo2,Mezzacapo3,Mezzacapo4,Mezzacapo5,Neuscamman2,AlAssam} for the ground state wave function of the system and 
variational Monte Carlo (VMC).\cite{AlAssam,Foulkes, Sandvik, Lou, Metropolis} In the EPS approach the lattice is covered with plaquettes
and the ground state wave-function is written in terms of the plaquette coefficients. Configurational weights are then optimized using a VMC algorithm. Very accurate results 
have been obtained for several frustrated and unfrustrated models within the EPS approach. The approach overcomes some of the limitations of two main numerical techniques,
density matrix renormalization group (DMRG) \cite{White} and quantum Monte Carlo (QMC),\cite{Ceperley} that have been successfully applied to study various quantum many-body systems. 
While DMRG gives very accurate results only in one dimension \cite{Liang} and QMC suffers from the sign problem for Fermi systems, the EPS and VMC approach can be applied to
systems of any spatial dimensionality and is sign problem free.
\\
\indent
The values of the plaquette coefficients that minimize the energy can be found using the stochastic minimization method.\cite{AlAssam, Sandvik, Lou} Due to the presence of
the statistical error in the stochastic algorithm it is difficult to obtain good estimates of the ground state energies for small system sizes. Having a larger number of parameters
allows the optimization method more freedom in finding the minimum energy state and the statistical error can be controlled by increasing the sample size. Therefore, instead
of comparing the EPS ansatz results with ED results for small clusters, we show that the EPS results for larger system sizes predict the same phases of the system as ED 
results predict for small clusters. In particular, we calculate within the EPS approach the density profiles and the density-functional fidelity\cite{Gu2} for a range of parameter values. This then
provides a confirmation of the validity of the EPS ansatz to describe the ground states of the system and a further confirmation of the existence
of the QAH state in the thermodynamic limit, as predicted by the mean-field theory calculations. 
\\
\indent
The manuscript is organized as follows. In Sec. \ref{sec:Model} we introduce the extended Hubbard model for spinless fermions on a honeycomb lattice. In Sec. \ref{sec:MF} we
review the mean-field phase diagram for the half-filled case. In Sec. \ref{sec:ED} we present results for small clusters with open boundary condition. The results for larger
system sizes obtained within the EPS and VMC approach are presented in Sec. \ref{sec:EPSandVMC}. In the final section Sec. \ref{sec:Conclusions} we draw our conclusions and 
discuss possible directions for future research.

\section{Model}
\label{sec:Model}
We consider the system of interacting spinless fermions on a honeycomb lattice. The system can be described by an extended Hubbard model
\begin{equation}\label{eq:Hubbard}
H =-t\sum_{\langle ij\rangle}\left(c_i^{\dagger}c_j+c_j^{\dagger}c_i\right)+V_1\sum_{\langle ij\rangle}n_in_j + V_2\sum_{\langle\langle ij\rangle\rangle}n_in_j,
\end{equation}
where $c_i$ ($c_i^{\dagger}$) are the fermion creation (annihilation) operators at site $i$, $n_i=c_i^\dagger c_i$ is the number operator at site $i$, $V_1$ is a 
repulsive nearest neighbor (NN) interaction and $V_2$ is a repulsive next to nearest neighbor (NNN) interaction. Here $\langle ij \rangle$ and $\langle\langle ij \rangle\rangle$
denote NN and NNN sites $i$ and $j$, respectively.
\\
\indent
The honeycomb lattice is a bipartite lattice, consisting of two triangular sublattices
further referred to as sublattices A and B. In the noninteracting limit ($V_1=V_2=0$) the system is in the semimetal (SM) phase with two inequivalent Fermi points. The NN interaction $V_1$ is the interaction between NN fermions on two different sublattices and favors charge density wave phase (CDW) with an
order parameter $\rho = (\langle c_{iA}^\dagger c_{iA}\rangle-\langle c_{iB}^\dagger c_{iB}\rangle)/2$. However, the NNN interaction $V_2$ between NN fermions on the same sublattice 
introduces frustration that can cause suppression of the CDW order and appearance of the topologically nontrivial states in the phase diagram of the system. 
\\
\indent
For the system at half filling 
predicted topologically nontrivial state is the QAH state\cite{Raghu,Weeks,Grushin} characterized by spontaneously broken time reversal symmetry and a nonlocal bond order parameter 
$\chi_{ij}=\langle c_i^{\dagger}c_j\rangle$. Properties of the SM-QAH critical point have been investigated in several previous studies\cite{Herbut2,Herbut3} and the SM-QAH quantum phase
transition was also found at weak interaction and in the presence of strain.\cite{Herbut1,Roy}
\\
\indent
The mean-field theory calculations also predict charge modulated phase \cite{Grushin} (charge density wave with reduced rotational symmetry)
for larger values of the NNN interaction $V_2\gtrsim 2.5 t$, and Kekul\'{e} phase characterized by an alternating bond strength.\cite{Weeks,Grushin} We review the mean-field phase diagram 
in the following section. 

\section{Mean-field phase diagram}
\label{sec:MF}
\begin{figure}[b!]
\caption{\label{fig:MF}Mean-field phase diagram for the half-filling case obtained in Ref.15. 
The various phases are discussed in the text. SM stands for semimetal, QAH for the quantum anomalous Hall phase, CMs for the
charge modulated phase, CDW for the charge density wave phase, and K for the Kekul\'e phase.
\\ 
} 
\includegraphics[width=0.75\columnwidth]{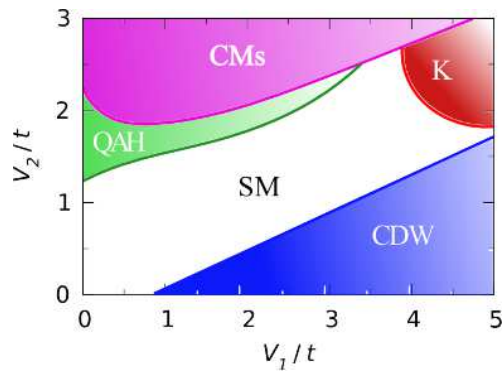}
\end{figure}

The mean-field phase diagram for the half-filled case, obtained by Grushin \emph{et al.}\cite{Grushin} is shown in Fig.\ref{fig:MF}. For the NN interaction $V_1=0$ and $V_2\leq V_{2c}^{(1)}$ the system is in the SM phase, and with increasing $V_2$ beyond the critical value $V_{2c}^{(1)}$ there is a continuous transition from the semimetal to the QAH 
phase and further a first-order transition to the charge modulated phase (CMs) at $V_2=V_{2c}^{(2)}$. The QAH phase is found for $1.5\lesssim V_2/t\lesssim 2.5$. In the SM phase that is connected to the noninteracting 
($V_1 = V_2 = 0$) limit of the Hamiltonian (\ref{eq:Hubbard}) there are two inequivalent Fermi points (Dirac points) where two energy bands intersect linearly.
In the vicinity of these Dirac points the electrons behave as relativistic two-dimensional massless Dirac fermions. 
\\
\indent
The QAH phase is an example of a topological phase that appears as a consequence of spontaneously broken time reversal symmetry. The QAH state has chiral edge states and a nonzero
Chern number $C=\pm 1$. The state is characterized by a complex bond order parameter $\chi_{ij}=\chi_{ji}^*=\langle c_i^\dagger c_j\rangle$ that corresponds to the complex hopping
term of the Haldane model\cite{Haldane} and breaks time reversal symmetry.
\\
\indent
The charge modulated phase (CMs) is characterized by a charge modulation within the same sublattice. The phase is a charge density wave with reduced rotational symmetry and it 
appears in the $V_2 \gtrsim V_1$ regime where it becomes energetically favorable to reduce the NNN energetic contribution $\propto V_2$ by paying the corresponding NN energy cost
$\propto V_1$. The CMs phase is a rather unconventional phase. The phase is a Mott insulator since it emerges from a large NNN interaction strength $V_2/t$. However, its
band gap is determined by the hopping integral $t$, that is a property of band insulators. 
\\
\indent
For $V_1 > 0$ a CDW phase and a Kekul\'{e} ordered phase were also found. The CDW phase with broken inversion symmetry appears for $V_1 > V_2$. The CDW order with
charge imbalance between the two different sublattices reduces the amount of NN interaction energy, and with increasing $V_2$ a Kekul\'{e} ordered phase appears at 
$V_2\sim V_1$. The Kekul\'{e} phase is characterized by an alternating bond strength, $Z_3$ order parameter and broken translational symmetry of the original honeycomb lattice that opens a gap in the
energy spectrum. 
\section{Exact diagonalization results for small clusters with open boundary condition}
\label{sec:ED}
\begin{figure}[b!]
\caption{\label{fig:C18}Illustration of the cluster of 18 sites studied in section \ref{sec:ED}. A an B labels correspond to two triangular sublattices of the honeycomb lattice. } 
\includegraphics[width=0.5\columnwidth]{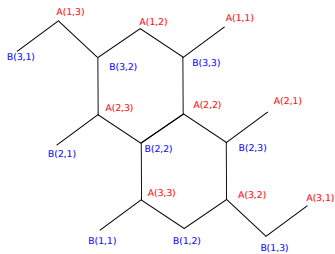}
\end{figure}
\begin{figure}[t!]
\caption{\label{fig:Delta3}The excitation gaps for the first three excited states as functions of the NNN interaction strength $\bar{V}_2=V_2/t$. Closing of the excitation gap at $\bar{V}_2=\bar{V}_{2c}^{(1)}$ marks the transition from SM to QAH state, and at
$\bar{V}_2=\bar{V}_{2c}^{(2)}$ the transition from QAH to CMs state.} 
\includegraphics[width=\columnwidth]{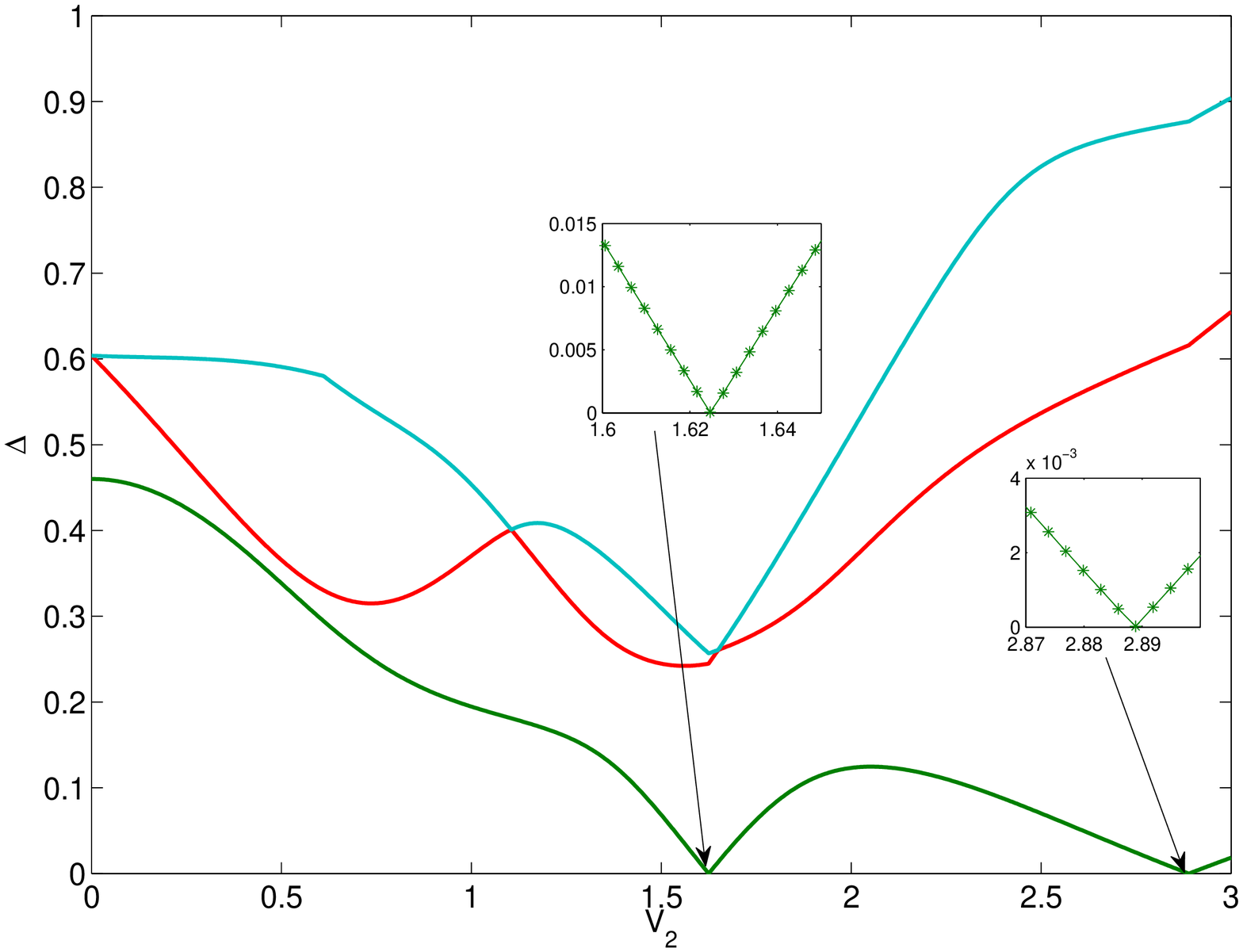}
\end{figure}
\begin{figure}[b!]
\caption{\label{fig:chi3x3}The fidelity susceptibility $\chi_F(\bar{V}_2,\delta \bar{V}_2)$ with $\delta\bar{V}_2 = 0.003$ as a function of the interaction strength $\bar{V}_2=V_2/t$ for the 18 sites cluster
shown in Fig. \ref{fig:C18}.}
\includegraphics[width=\columnwidth]{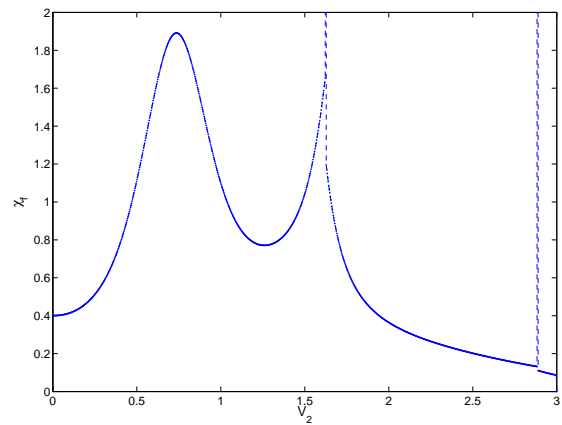}
\end{figure}

Using exact diagonalization (ED) method, we first examine signatures of topological phase transitions for small clusters with open boundary condition. The key signature of 
the topological phase transition is the existence of a topologically protected level crossing that is robust and defines a topological phase transition even in a finite size
system.\cite{Varney} For the interaction-driven topological transition the level crossing corresponds to the closing of the excitation gap. Accordingly, a topological transition
in an interacting system is marked by the closing of the excitation gap. Choice of the boundary condition that can be used to detect topological order in finite size systems depends
on spatial symmetries of the system. 
\\
\indent
The level crossing can also be observed in the fidelity metric.\cite{Varney, Wang, Gu, Abasto, Eriksson, Varney2, Zanardi, Venuti, Rigol, Varney3} If we denote by $|\psi_0(\beta)\rangle$ and $|\psi_0(\beta+\delta\beta)\rangle$
two ground states corresponding to slightly different values of the relevant parameter $\beta$, the fidelity between the two ground states is equivalent to the modulus of the overlap between the two states:
\begin{equation}\label{eq:Fidelity}
F(\beta,\beta+\delta\beta)=|\langle\psi_0(\beta+\delta\beta)|\psi_0(\beta)\rangle|.
\end{equation}
The equation (\ref{eq:Fidelity}) can be rewritten as 
\begin{equation}\label{eq:chiF}
F=1-\frac{(\delta\beta)^2}{2}\chi_F+...,
\end{equation}
where $\chi_F$ is the fidelity susceptibility,
\begin{equation}\label{eq:chiF2}
\chi_F(\beta)= \lim_{\delta\beta\rightarrow 0}\frac{-2\ln F}{\delta\beta^2}=-\frac{\partial^2 F}{\partial(\delta\beta)^2}. 
\end{equation}
The overlap measures similarity between two states, it gives unity if two states are the same and zero if the states are orthogonal. At the point of level-crossing between 
two orthogonal states the fidelity shows a very sharp drop that corresponds to a singular peak in the fidelity susceptibility. The topological transition can then be characterized by a singular peak in the fidelity susceptibility. 
\begin{figure}[t!]
\caption{\label{fig:E3}The ground state and the first two excited states energies as functions of the NNN interaction strength $\bar{V}_2=V_2/t$. The insets show a closeup view of the level crossing at 
$\bar{V}_2=\bar{V}_{2c}^{(1)}$ and $\bar{V}_2=\bar{V}_{2c}^{(2)}$. The level crossing at $\bar{V}_2=\bar{V}_{2c}^{(1)}$ marks the second order transition from SM to QAH state, where the value of the topological index (Chern number)
changes from zero to one. The level crossing at $\bar{V}_2=\bar{V}_{2c}^{(2)}$ corresponds to the first order phase transition from the QAH to CMs phase accompanied with the change of the value of the Chern number from one to zero.} 
\includegraphics[width=\columnwidth]{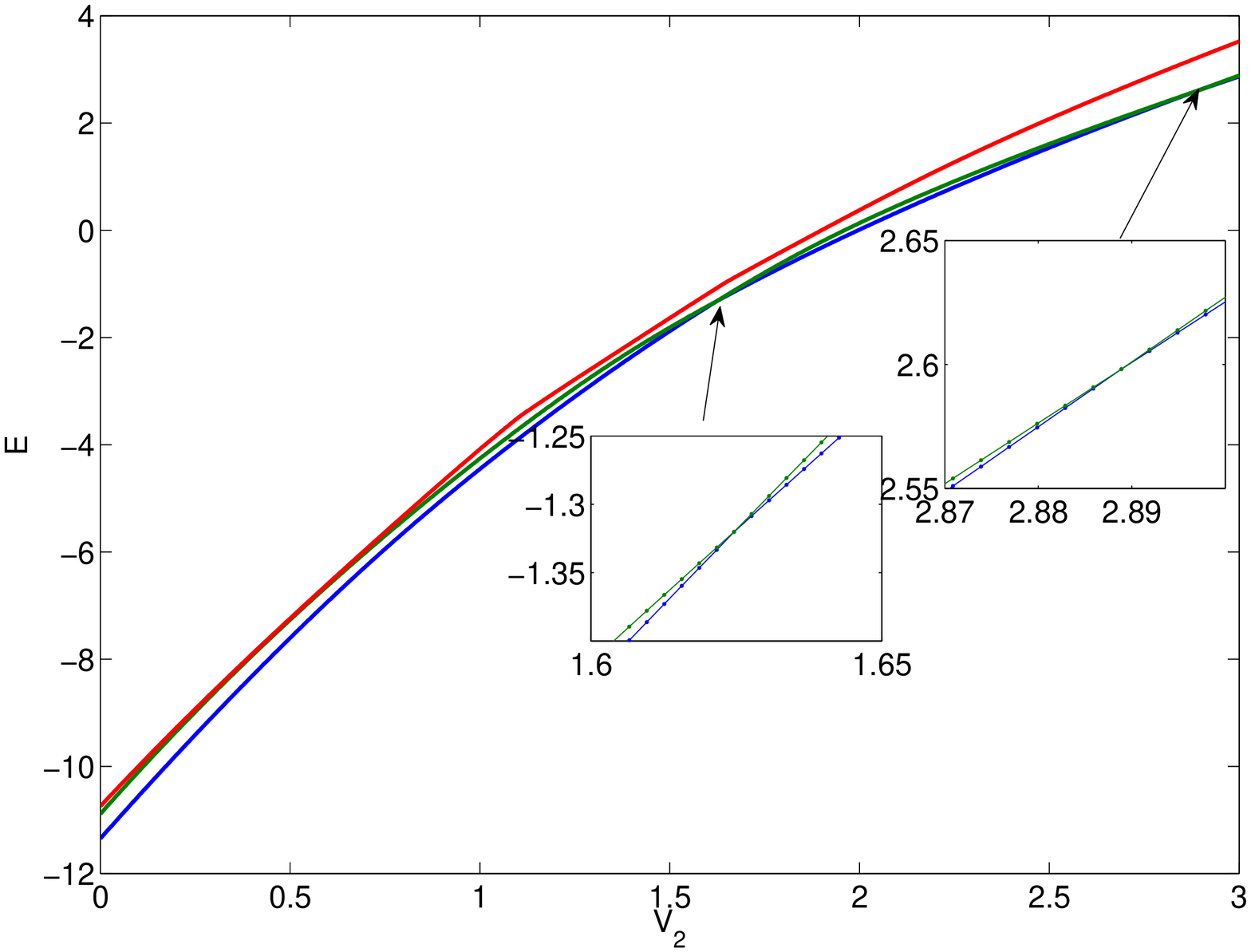}
\end{figure}
\\
\indent
Since we are primarily interested in examining stability of the QAH phase that is, according to the mean-field theory calculations, stabilized by the NNN interaction 
$V_2$, we focus on the case $V_1=0$ for which the QAH phase is most extended in ($V_1$, $V_2$) parameter space. 
In the ED calculations the finite size effects can exceed the energy scale of the many-body energy gap of an incompressible ground state, so that the incompressibility 
can not be recognized. For the topologically nontrivial insulating states choice of open boundary conditions, instead of periodic boundary conditions, changes the energy
spectrum of the system by introducing the edge states within the bulk energy gap. In some cases this allows identification of the topological insulating states from the 
ED calculations, even when such states can not be identified from the ED calculations with periodic boundary conditions. We therefore study properties of the system for 
several small system sizes and with open boundary condition. 
\\
\indent
We present the results for the cluster of 18 sites (two $3\times 3$ triangular sublattices) illustrated in Fig.\ref{fig:C18}, for which signatures 
of the topological phase transitions predicted by the mean-field theory calculations can be clearly seen. For smaller systems sizes and different aspect ratios (clusters with
two $2\times 2$, $2\times 3$, $2\times 4$ and $3\times 4$ triangular sublattices) we do not observe the level crossings. That is consistent
with the previous findings\cite{Varney,Varney2} that the topological transitions can be detected by the ED calculations only if one studies clusters with a reciprocal space 
that contains Dirac points. 
\\
\indent
The energies of the ground state and first two excited states as functions of the NNN interaction strength $\bar{V}_2=V_2/t$ are shown in Fig.\ref{fig:E3}. The level crossing at 
$\bar{V}_2=\bar{V}_{2c}^{(1)}$ marks the second order transition from SM to QAH state, where the value of the topological index (Chern number)
changes from zero to one. The level crossing at $\bar{V}_2=\bar{V}_{2c}^{(2)}$ corresponds to the first order phase transition from the QAH to CMs phase accompanied with the 
change of the value of the Chern number from one to zero. The level crossings can also be clearly seen in Fig.\ref{fig:Delta3} that shows the excitation gaps $\Delta^{(n)}=E_n-E_0$ for the first three excited states ($n=1,2$ and $3$). 
In addition, the topological transitions are characterized by singular points in the fidelity susceptibility shown in Fig.\ref{fig:chi3x3}. The fidelity and fidelity susceptibility are defined by equations
(\ref{eq:Fidelity})-(\ref{eq:chiF2}), where here the relevant parameter $\beta = \bar{V}_2$.
\begin{figure}[t!]
\caption{\label{fig:EdgeStates3x3}The density profile of a hole created out of the ground state ($\Delta n_j$) at half-filling for the 18 sites cluster
shown in Fig.\ref{fig:C18} and for the interaction strength a) $\bar{V}_2=0$, b) $\bar{V}_2=1.5$, c) $\bar{V}_2=2$ and d) $\bar{V}_2=3$ . 
Here the radius of each filled circle is proportional to the magnitude of $|\Delta n_j|$ and empty circles denote lattice sites where $\Delta n_j<0$.}
\includegraphics[width=\columnwidth]{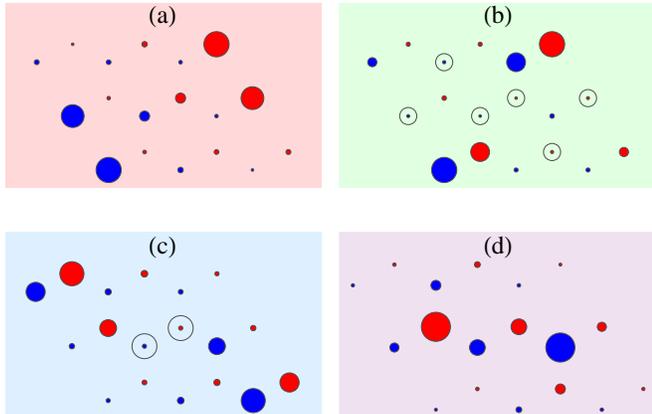}
\end{figure}
\\
\indent
We further examine the existence of the edge states in the $\bar{V}_{2c}^{(1)}<\bar{V}_2<\bar{V}_{2c}^{(2)}$ regime by considering the density profile of a hole created out of the
ground state.\cite{Li,Varney4} The energy of single particle (hole) excitation is defined as $\Delta E_p=E_0^{N+1}-E_0^{N}$ ($\Delta E_h=E_0^N-E_0^{N-1}$), where $E_0^N$ is the ground state energy of the system with $N$ 
fermions. The density profile of a hole created out of the ground state is 
\begin{equation}\label{eq:DeltaN}
\Delta n_j=\langle \psi_0^{N}|n_j|\psi_0^{N}\rangle-\langle\psi_0^{N-1}|n_j|\psi_0^{N-1}\rangle,
\end{equation}
where $|\psi_0^N\rangle$ is the ground state wave-function with $N$ fermions. In the presence of the in-gap edge states the density profile should be localized on the edges. The density profile for several values of the NNN interaction strength $V_2/t$ 
is shown in Fig.\ref{fig:EdgeStates3x3}. The results confirm that for the value of $V_2/t=2$, when the system is in the QAH regime, the density profile is indeed localized 
on the edges, while for the values of $V_2/t=0,1.5$ and $3$, corresponding to SM and CMs regimes, the density profile is delocalized. Note that for $V_2/t=2$ the density profile
$\Delta n_j<0$ at the sites of the central plaquette and $>0$ at the sites of the plaquettes at the edges of the cluster. 
\\
\indent
We also note that for a finite cluster in the QAH regime the ground state obtained from ED will be a superposition of two spontaneously generated QAH states with opposite
chirality (Chern number $C=\pm1$). More precisely, the QAH state that we find is a linear combination of the two Haldane states with C=+1 and C=-1 which is odd under 
time-reversal. This can be seen by considering the time-reversal symmetry of the system. 
\\
\indent
For the spinless fermions the (antiunitary) time-reversal (TR) symmetry operator $T=\sigma_1 K$, where $\sigma_1$ denotes Pauli matrix and K complex conjugation. 
The TR symmetry operator $T$ is an exact symmetry operator of the Hamiltonian (\ref{eq:Hubbard}), and therefore it commutes with the Hamiltonian. In the basis of the 
two states that are related by the TR symmetry, $|\psi_+\rangle$ and $|\psi_-\rangle=T|\psi_+\rangle$, the Hamiltonian then has to be proportional to 
the unit matrix and to $\sigma_1$. However, although the system has TR symmetry the antiunitary TR symmetry operator squares to one ($T^2=+1$), and as such does not
imply Kramers' degeneracy. Accordingly, there is no two-fold degeneracy of the QAH phase in the finite size system we are studying. 
\\
\indent 
The symmetry associated with the two states that cross is the TR symmetry. The QAH state has a different quantum number for the TR symmetry operator than SM and CMs 
phases. The QAH ground state that we find is the linear combination of the two TR symmetry breaking states which is odd under TR, whereas the SM and CMs phases are even under TR.
Since the Hamiltonian can not have matrix elements between states that are odd and even under TR the level crossings are allowed. 
\\
\indent
In summary, our ED results show clear signs of an interaction-driven Chern insulator (QAH) phase, contrary to recent ED studies for small clusters with periodic boundary
condition.\cite{Daghofer, Martinez} This confirms importance of the boundary condition and the symmetry of the cluster to detect topological order in finite size systems. 
In the following section we study the ground state properties of the system for larger system sizes to further confirm the existence of the QAH phase in the thermodynamic limit.
\section{The EPS ansatz and variational Monte Carlo algorithm}
\label{sec:EPSandVMC}
Studying larger system sizes with ED method is not possible due to the rapid increase in the size of the Hilbert space with the increase of the size of the system. 
However, recently there has been much success in using tensor network methods to numerically simulate variety of strongly correlated models. In this section we 
examine applicability of the entangled-plaquette state (EPS) ansatz\cite{Changlani,Mezzacapo1,Neuscamman,Mezzacapo2,Mezzacapo3,Mezzacapo4,Mezzacapo5,Neuscamman2,AlAssam} 
to describe the ground states of the system. Within this approach the lattice is covered with plaquettes
and the ground state wave-function is written in terms of the plaquette coefficients. Configurational weights can then be optimized using a variational Monte Carlo (VMC)
algorithm. This then allows to study the ground state properties of the system for larger system sizes and further confirmation of the existence of the QAH phase in the thermodynamic limit. 
\begin{figure}[b!]
\caption{\label{fig:Plaquettes}Illustration of the 8 sites plaquette correlators used in the calculation of the ground state properties of the system for larger system sizes. Here the underlying triangular
lattice has two-site basis, where the basis sites are labeled by A and B.}
\includegraphics[width=0.75\columnwidth]{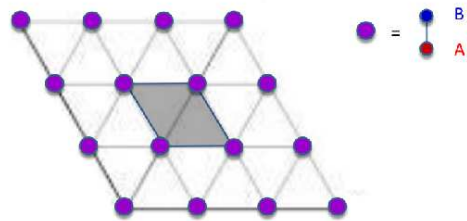}
\end{figure}
\\
\indent
For a lattice with $l$ sites an arbitrary quantum many-body wave function can be written as
\begin{eqnarray}\label{eq:wavefunction}
|\psi\rangle&=&\sum_{n_1,n_2,...,n_{l}}W_{n_1,n_2,...,n_{l}}|n_1,n_2,...,n_{l}\rangle\\ \nonumber
&=&\sum_{\mathbf{n}}W_{\mathbf{n}}|\mathbf{n}\rangle,
\end{eqnarray}
where $\mathbf{n}$ denotes the vector of occupancies ($n_i\in\{0,1\}$ in a fermion system) and $W_{\mathbf{n}}$ is the amplitude or weight of a given configuration $\mathbf{n}$.
In the EPS description, also called correlator product state (CPS) description, the weight $W(\mathbf{n})$ is expressed as a product of the plaquette coefficients over the lattice.
A correlator is an operator that is diagonal in the lattice basis and can be chosen to act on an arbitrary number of sites. Such a general correlator for a plaquette $p$ with $l_p$ sites can be written as 
\begin{equation}\label{eq:correlator}
\hat{c}_p= \sum_{\mathbf{n}_p}c_p^{\mathbf{n}_p}\hat{P}_{\mathbf{n}_p},
\end{equation}
where $\mathbf{n}_p=\{n_{p1},n_{p2},...,n_{pl_p}\}$ is the occupancy vector of the sites of the plaquette $p$ and $\hat{P}_{\mathbf{n}_p}$ is the projection operator
\begin{equation}\label{eq:P_Operator}
\hat{P}_{\mathbf{n}_p}=|\mathbf{n}_p\rangle\langle\mathbf{n}_p|.
\end{equation}
The EPS (CPS) is then obtained by applying a product of correlators to a reference wavefunction $|\Phi\rangle$
\begin{equation}\label{eq:EPS_wf}
|\psi\rangle_{EPS}=\hat{C}|\Phi\rangle=\prod_p \hat{c}_p|\Phi\rangle.
\end{equation}
For a fermionic system the Jastrow-Slater CPS fermion wave function is obtained by applying a set of correlators (Jastrow factors) to a Slater determinant reference of orbitals \cite{Neuscamman}
\begin{equation}\label{eq:determinant}
|\Phi\rangle=\det|\phi_1\phi_2....\phi_k|
\end{equation}
and the wave function amplitudes $W_{\mathbf{n}}$ in the equation (\ref{eq:wavefunction}) are in this case given by 
\begin{equation}\label{eq:EPS_W}
W_{n_1,n_2,...,n_l}=\prod_p c^{\mathbf{n}_p}_p\times \det |\phi_1(r_1)\phi_2(r_2)...\phi_l(r_k)|,
\end{equation}
where $r_1,r_2,...,r_k$ label the positions of $k$ occupied sites in the occupancy vector $|n_1,n_2,...,n_l\rangle$. 
\\
\indent
Here the Slater determinant (\ref{eq:determinant}) orbitals can be obtained by solving the noninteracting limit of the Hamiltonian (\ref{eq:Hubbard}) with $V_1=V_2=0$. By applying
a product of correlators to a Slater determinant reference wave function the correlations between sites are introduced into the wave function EPS ansatz and are explicitly
encoded in correlator building blocks. The wave function of the original system, expressed as the product of the wave functions for sub-blocks (plaquettes) gives reasonable estimates 
of the ground state energy and short-range correlations.
\begin{figure}[t!]
\caption{\label{fig:E0_EPS}The ground state energy at half filling as a function of the NNN interaction strength $V_2/t$ ($V_1=0$) for the 128 sites cluster of the same shape as the cluster in Fig.\ref{fig:Plaquettes}
and with open boundary condition, obtained using the EPS ansatz with 8 sites plaquettes (Fig.\ref{fig:Plaquettes}) and stochastic minimization method. 
Statistical errors are smaller than the symbol size.}
\includegraphics[width=0.75\columnwidth]{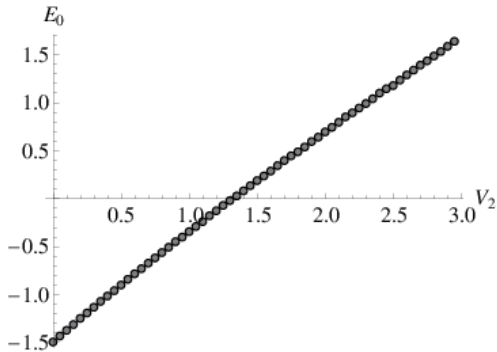}
\end{figure}
\begin{figure}[b!]
\caption{\label{fig:DFFS}Density-functional fidelity susceptibility $\chi_{F_n}(\bar{V}_2,\delta \bar{V}_2)$ with $\delta\bar{V}_2 = 0.05$ as a function of the interaction strength $\bar{V}_2=V_2/t$ ($V_1=0$) for the 128 sites cluster of the same shape as the cluster in Fig.\ref{fig:Plaquettes}
and with open boundary condition. Statistical errors are smaller than the symbol size. SM to QAH phase transition (a) and QAH to CMs phase transition (b) are characterized by a singular peak (discontinuity) in $\chi_{F_n}(\bar{V}_2,\delta \bar{V}_2)$. }
\includegraphics[width=0.75\columnwidth]{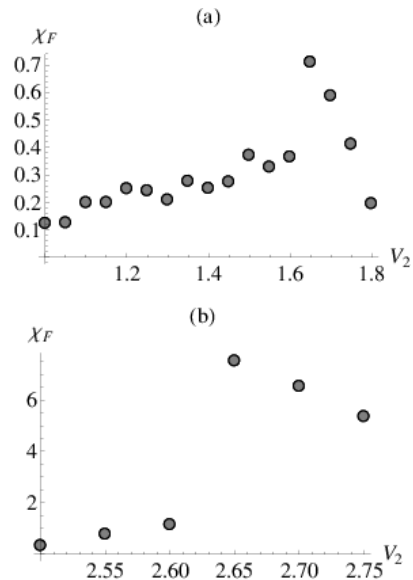}
\end{figure}
\\
\indent
The estimates improve with the increase of the plaquette size and also, in general, the greater overlap between the plaquettes 
gives more accurate description of the ground state. By increasing the number of sites covered by correlators (larger plaquette size) the EPS becomes an exact family of states.
However, in many cases even correlators with a small number of parameters can describe the qualitative behavior of large systems. Here we choose 8 sites plaquettes illustrated in
Fig.\ref{fig:Plaquettes} and consider as an example a cluster of $128$ sites (two $8\times8$ triangular sublattices) of the same shape as the cluster in Fig.\ref{fig:Plaquettes}
and with open boundary condition. 
\\
\indent
Expectation values of the energy and other operators can be obtained by using Monte Carlo importance sampling.\cite{AlAssam, Foulkes,Sandvik, Lou} In variational Monte Carlo 
(VMC) the energy $E$ is written as 
\begin{equation}\label{eq:E_VMC}
E=\langle \psi|H|\psi\rangle=\frac{\sum_{\mathbf{n},\mathbf{n}'}W^*(\mathbf{n}')\langle \mathbf{n}'|H|\mathbf{n}\rangle W(\mathbf{n})}{\sum_{\mathbf{n}}|W(\mathbf{n})|^2},
\end{equation}

where we assume that the wavefunction $|\psi\rangle$ is not normalized. The energy can then be further rewritten as 
\begin{equation}\label{eq:E_VMC2}
E=\sum_{\textbf{n}}P(\textbf{n})E(\textbf{n}),
\end{equation}
where $E(\textbf{n})$ is the local energy 
\begin{equation}\label{eq:En}
E(\textbf{n})=\sum_{\textbf{n}'}\frac{W^*(\textbf{n}')}{W^*(\textbf{n})}\langle\textbf{n}'|H|\textbf{n}\rangle
\end{equation}
and the probability $P(\textbf{n})$ is given by 
\begin{equation}\label{eq:Probability}
P(\textbf{n})=\frac{|W(\textbf{n})|^2}{\sum_{\textbf{n}}|W(\textbf{n})|^2}. 
\end{equation}
The expectation value of any operator $\hat{O}$ can be written in the same form by replacing the Hamiltonian $H$ with the operator $\hat{O}$ in the previous expressions. According to the 
variational principle minimization of the expression (\ref{eq:E_VMC2}) with respect to the weights gives an upper bound of the ground state energy. The probability of a 
given configuration $P(\textbf{n})$ is never explicitly calculated from the equation (\ref{eq:Probability}). Instead, the Metropolis algorithm \cite{Metropolis} is used to sample 
the probability distribution and to efficiently compute the overall energy as an average of the sampled local energies.
\\
\indent
The plaquette coefficients that minimize the energy can be found by using the stochastic minimization method \cite{AlAssam, Foulkes, Sandvik, Lou} which requires only the first 
derivatives of the energy with respect to the plaquette coefficients given by
\begin{equation}\label{eq:partial_E}
\frac{\partial E}{\partial c_p^{\mathbf{n}_p}} =2\sum_{\mathbf{n}}\{ P(\mathbf{n})\Delta_p^{\mathbf{n}_p}[E(\mathbf{n})-\sum_{\mathbf{n}'}P(\mathbf{n}')E(\mathbf{n}')]\},
\end{equation}
where the wavefunction $|\psi\rangle$ in the equation (\ref{eq:E_VMC}) is approximated by the EPS wavefunction (\ref{eq:EPS_wf}) and 
\begin{equation}\label{eq:Delta_EPS}
\Delta_p^{\mathbf{n}_p}= \frac{1}{W(\mathbf{n})}\frac{\partial W(\mathbf{n})}{\partial c_p^{\mathbf{n}_p}}=\frac{b_p}{c_p^{\mathbf{n}_p}}
\end{equation}
with $W(\mathbf{n})$ given by the equation (\ref{eq:EPS_W}). Here $b_{p}$ denotes the number of times the plaquette coefficient $c_p^{\mathbf{n}_p}$ appears in
the product (\ref{eq:EPS_W}) for the amplitude $W(\textbf{n})$ for the configuration $\textbf{n}$. If each correlator is used only once $b_{p}=1$.

To calculate the ground state energy and the ground state expectation value of any operator $\hat{O}$, we follow the variational algorithm used in several previous studies
of various strongly correlated models.\cite{AlAssam, Sandvik, Lou}
\begin{figure}[t!]
\caption{\label{fig:n_1.8} The ground state density distribution for 128 sites cluster 
with open boundary condition at $V_2/t=1.8$ ($V_1=0$), when the system is in the QAH phase. 
Here the radius of each circle is proportional to the magnitude of the density. 
}
\includegraphics[width=0.75\columnwidth]{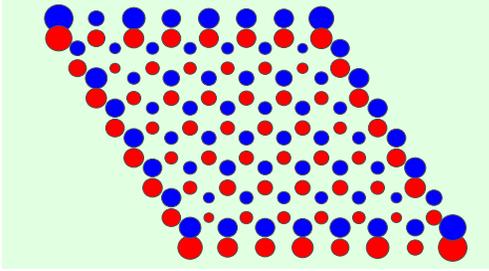}
\end{figure}
\\
\indent
For a given set of plaquette coefficients the energy (\ref{eq:E_VMC2}) and its first derivative (\ref{eq:partial_E}) 
can be efficiently computed using the Metropolis algorithm. In our calculation the total number of fermions $N$ is fixed. 
We start from a randomly chosen initial configuration $|\textbf{n}\rangle=|n_1,n_2,...,n_l\rangle$ with $\sum_{i=1}^l n_i= N$, and then generate via the Metropolis algorithm a large set of new configurations by exchanging the occupancy numbers $n_i$ and $n_j$ at two neighboring 
sites $i$ and $j$. Starting from the configuration $|\mathbf{n}\rangle$, the acceptance probability for a new configuration $|\mathbf{n}'\rangle$ is given by
\begin{equation}\label{eq:P_A}
P_A=\min\left[\frac{|W(\mathbf{n}')|^2}{|W(\mathbf{n})|^2},1\right].
\end{equation}
The overall energy can then be efficiently computed as an average of the local energies sampled by a Markov chain in the Metropolis algorithm.
The first derivative can be calculated equivalently from the same sample.  
\\
\indent
It is also important to note that there is no need to compute the wave function normalization at any point in the calculation. The most time-consuming step in the calculation 
is the evaluation of the fraction $W(\textbf{n}')/W(\textbf{n})$. However, the configuration weights $W(\textbf{n})$ are given by a simple product of numbers, and to calculate 
the fraction $W(\textbf{n}')/W(\textbf{n})$ it is necessary to calculate only products of the plaquette coefficients that represent sites $\left\{i\right\}$ where the occupancy numbers $n_{\left\{i\right\}}$ have changed
to $n'_{\left\{i\right\}}\neq n_{\left\{i\right\}}$. Also, if the Hamiltonian is local, there are only a few nonzero matrix elements $\langle\textbf{n}|H|\textbf{n}'\rangle$ in the expression (\ref{eq:En}) for the local energy,
and only a few fraction terms $W(\textbf{n}')/W(\textbf{n})$ need to be calculated for each contribution to the energy. 
\begin{figure}[b!]
\caption{\label{fig:n_2.8}
The ground state density distribution for 128 sites cluster 
with open boundary condition at $V_2/t=2.8$ ($V_1=0$), when the system is in the CMs phase.
Here the radius of each circle is proportional to the magnitude of the density. 
}
\includegraphics[width=0.75\columnwidth]{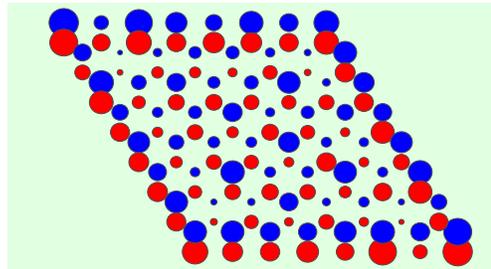}
\end{figure}
\\
\indent
The values of the plaquette coefficients that minimize the energy can be found using the stochastic minimization method.\cite{AlAssam, Sandvik, Lou} The steps of the variational
algorithm to calculate the ground state energy are: ($1$) start from the randomly chosen complex values for the plaquette coefficients, ($2$) evaluate the energy and its gradient 
vector, ($3$) update all the plaquette coefficients $c_p^{\mathbf{n}_p}$ according to 
\begin{equation}\label{eq:update_c}
c_p^{\mathbf{n}_p}\rightarrow c_p^{\mathbf{n}_p}-r\delta (k)\cdot \text{sign}\left(\frac{\partial E}{\partial c_p^{\mathbf{n}_p}}\right)^*,
\end{equation}
and ($4$) iterate from (2) until convergence of the energy is reached. Here $r$ is a random number between 0 and 1, and $\delta(k)$ is the step size for a given iteration $k$. 
\\
\indent
The energy and its derivative are estimated from $3\times F(k)\times l$ sampled values in each iteration $k$, where $l$ is the number of lattice sites and $F(k)$ is called the number of sweeps
per sample. In a given sweep each lattice site is visited sequentially and a move $\textbf{n} \rightarrow \textbf{n}'$ to a new configuration is proposed by exchanging the 
occupancy numbers $n_i$ and $n_j$ of two neighboring sites $i$ and $j$ (3 NN for each lattice site). Also, to achieve convergence and reach the optimal energy value it is important to 
carefully tune the gradient step $\delta(k)$. For each iteration $k$, the number of sweeps $F$ is increased linearly, $F=F_0k$, and the procedure of evaluating the energy 
and updating the coefficients is repeated $G=G_0k$ times. The step size is gradually reduced per iteration. Here we use a geometric form $\delta =\delta_0Q^k$ with $Q=0.9$. 
\\
\indent
The number of sweeps per iteration is increased because the derivatives become smaller as the energy minimum is approached and require more sampling in order not to be
dominated by noise. Increasing G effectively corresponds to a slower cooling rate. Here we take $F_0=10$, $G_0=10$ and $Q=0.9$. The initial minimization routine is performed
with $\delta_0=0.05$ for 50 iterations. The resulting plaquette coefficients are then used as a starting point for a new run of 50 iterations with $\delta_0=0.005$. After 
the minimization is complete the expectation values are calculated by repeating the procedure for a single iteration with zero step size and large $F$ and $G$ to obtain 
more accurate estimates of the expectation values. The results for the ground state energy as a function of the NNN interaction strength $V_2/t$ ($V_1=0$) for the 128 sites cluster with open boundary 
condition are shown in Fig. \ref{fig:E0_EPS}. 
\\
\indent
To confirm the topological phase transitions found for the smaller system sizes we further calculate the density-functional fidelity and its susceptibility\cite{Gu2} for the 128 sites cluster.
The density-functional fidelity measures the similarity between density distributions of two ground states in parameter space. As explained in Sec.\ref{sec:ED} the topological
transition can be characterized by a singular peak in the fidelity susceptibility (\ref{eq:chiF2}). However, according to the Hohenberg-Kohn theorems\cite{Hohenberg-Kohn} the ground state properties of a quantum many-body system are uniquely determined by the density distribution
$n_{\textbf{r}}$ that minimizes the ground-state energy functional $E_0\left[n_{\textbf{r}}\right]$. Therefore the most relevant information about the ground state of the system is captured by the
density distribution and any change in the structure of the wavefunction corresponds to a change of the density distribution. Accordingly the topological phase transition 
can be found by calculating the similarity between density distributions and is characterized by a singular peak in the density-functional fidelity susceptibility.
Moreover, the density-functional fidelity can be easily measured in experiments and provides a strategy to study quantum critical phenomena both theoretically and experimentally. 
\\
\indent
The density distribution can be obtained as
\begin{equation}\label{eq:density}
n_{\textbf{r}} = \langle \Psi_0(\beta)|\hat{n}_{\textbf{r}}|\Psi_0(\beta)\rangle,
\end{equation}
where $\textbf{r}=(x,y)$ denotes sites of the honeycomb lattice, $\hat{n}_{\textbf{r}}=c^\dagger_{\textbf{r}}c_{\textbf{r}}$ is the number operator at site $\textbf{r}$, and $\beta$ is the relevant
parameter in the Hamiltonian (here $\beta=V_2/t$). The density-functional fidelity for two ground states at $\beta$ and $\beta'$ is given by\cite{Gu2}
\begin{equation}\label{eq:DFF}
F_n(\beta,\beta')=\text{tr}\sqrt{n(\beta)n(\beta')},
\end{equation}
where 
\begin{equation}\label{eq:DFF2}
n=\sum_{\textbf{r}}n_{\textbf{r}}|\textbf{r}\rangle\langle\textbf{r}|. 
\end{equation}
The equation (\ref{eq:DFF}) can be rewritten as
\begin{equation}\label{eq:DFF3}
F_n(\beta,\beta+\delta\beta)= 1-\frac{(\delta\beta)^2}{2}\chi_{F_n}+...
\end{equation}
where $\delta\beta=\beta-\beta'$ and the density-functional fidelity susceptibility $\chi_{F_n}$ has the form
\begin{equation}\label{eq:DFF4}
\chi_{F_n}=\sum_{\textbf{r}}\frac{1}{4n_{\textbf{r}}}\left(\frac{\partial n_{\textbf{r}}}{\partial \beta}\right)^2
\end{equation}
Density-functional fidelity is a functional of $n_{\textbf{r}}$ and $\partial n_{\textbf{r}}/\partial \beta$ that both maximize the density-functional fidelity susceptibility (\ref{eq:DFF4})
at the critical point (for example if the density in a certain region vanishes rapidly). 
\\
\indent
The topological phase transitions from SM to QAH phase and from QAH phase to CMs phase for 128 sites cluster, characterized by singular points in the density-functional fidelity susceptibility, can be clearly seen in Fig.\ref{fig:DFFS}. Also,
a charge modulation within the same sublattice in the CMs phase can be clearly seen in the density distribution as shown in Fig.\ref{fig:n_2.8}, while the charge modulation
is absent in the QAH phase (Fig.\ref{fig:n_1.8}). Our results indicate that for the 128 sites cluster and at $V_1=0$ the QAH phase is found for $1.6\lesssim V_2/t\lesssim 2.6$, that is very close to the 
mean-field theory predictions for the system in the thermodynamic limit and to the ED result obtained for a much smaller system size (18 site cluster). 
\\
\indent
We also note that in the thermodynamic limit the two lowest energy states, 
that are odd and even under TR, become exactly degenerate in the QAH regime. Consequantly there are two degenerate QAH ground states with oposite chirality (Chern number $C=\pm 1$) in the 
thermodynamic limit. That also leads to the disappearance of the cusp in the energy as a function of $V_2/t$ at the SM-QAH phase transition, and confirms that the SM-QAH transition is a continuous phase
transition as predicted by the mean-field theory calculations. 
\\
\indent
In summary, our results obtained using the EPS ansatz for the ground state wavefunction and VMC indicate existence of the QAH state for larger system sizes and provide
further confirmation of the presence of the QAH phase in the thermodynamic limit, as predicted by several mean-field theory calculations.\cite{Raghu, Weeks, Grushin} This is in contrast to the previous
results \cite{Daghofer, Martinez} obtained using ED method for small clusters with periodic boundary condition and cluster perturbation theory. \cite{Daghofer} 

\section{Conclusions}\label{sec:Conclusions}
We have studied the system of interacting spinless fermions on a honeycomb lattice. Using exact diagonalization method for small system sizes, and entangled-plaquette state
ansatz and variational Monte Carlo method for larger system sizes, we find evidence for the existence of the interaction generated quantum Hall state previously predicted by
several mean-field theory calculations.\cite{Raghu, Weeks, Grushin} In particular, we presented results for 18 sites cluster and 128 sites cluster with open boundary condition. We find clear signs of
the predicted topological phase transitions (from semimetal to interaction-generated quantum Hall state and from quantum Hall state to charge modulated state) in the 
fidelity metric and also demonstrate the appearance of the sublattice charge modulation in the charge modulated phase by calculating the density profile for the 128 sites cluster.
The quantum Hall state was not found in the previous exact diagonalization studies of small clusters with periodic boundary condition.\cite{Daghofer, Martinez} Our results
thus indicate importance of the boundary condition and the symmetry of the cluster to detect topological order in finite size systems. The results also confirm validity 
of the Jastrow-Slater entangled-plaquette state (correlator-product state) ansatz to describe the ground states of the system. 
\\
\indent
Further work is necessary for more direct comparison of the entangled-plaquette state ansatz results with the exact diagonalization results for small clusters. As mentioned before, a disadvantage
of the variational Monte Carlo stochastic algorithm is the presence of the statistical error which can be controlled by increasing the system size. To obtain the results that can
be directly compared with the exact diagonalization results for small clusters it is therefore necessary to use nonstochastic algorithm\cite{Neuscamman} to evaluate the energy and to 
optimize the amplitudes of the entangled-plaquette state wavefunction. Also, further insights could be obtained by considering the results using different types of 
correlators (plaquettes of different shapes and sizes). Additional directions of future research are to study the ground states of the system at nonzero nearest neighbor 
interaction strength $V_1$ and possible topological phases of the system away from half-filling that are predicted by mean-field theory calculations.\cite{Grushin,Castro} 
We hope that the results presented in this manuscript will motivate further studies of the extended Hubbard model for fermions on a honeycomb lattice and other models for which
the interaction-generated topologically nontrivial phases have been predicted.

\begin{acknowledgments}

We thank Fakher F. Assaad, Derek K. K. Lee and Chris Hooley for very helpful suggestions and discussions.
We also thank Andrew G. Green for very useful suggestions and for carefully reading the manuscript. This
work was funded by the EPSRC under grant code EP/K02163X/1. Nicholas
Chancellor was funded by Lockheed Martin Corporation at the time this
work was carried out. Igor F. Herbut is supported by the NSERC of Canada.

\end{acknowledgments}

\end{document}